\documentclass[prl,
twocolumn,
 amsmath,amssymb,
 aps,
]{revtex4-2}

\usepackage{graphicx}
\usepackage{dcolumn}
\usepackage{bm}
\usepackage[utf8]{inputenc}
\usepackage{todonotes}
\usepackage{comment}

\newcommand{\ri}{\mathrm{i}}

\makeatletter
\def\moverlay{\mathpalette\mov@rlay}
\def\mov@rlay#1#2{\leavevmode\vtop{%
   \baselineskip\z@skip \lineskiplimit-\maxdimen
   \ialign{\hfil$\m@th#1##$\hfil\cr#2\crcr}}}
\newcommand{\charfusion}[3][\mathord]{
    #1{\ifx#1\mathop\vphantom{#2}\fi
        \mathpalette\mov@rlay{#2\cr#3}
      }
    \ifx#1\mathop\expandafter\displaylimits\fi}
\makeatother

\begin{document}

\title{Exact Non-Local Hydrodynamics Predict Rarefaction Effects}

\author{Florian Kogelbauer}
\email{floriank@ethz.ch}\thanks{Corresponding Author}
\affiliation{Department of Mechanical and Process Engineering, ETH Z{u}rich, 8092 Z{u}rich, Switzerland}

\author{Ilya Karlin}
\email{ikarlin@ethz.ch}
\affiliation{Department of Mechanical and Process Engineering, ETH Z{u}rich, 8092 Z{u}rich, Switzerland}

\date{\today}

\begin{abstract}
We combine the theory of slow spectral closure for linearized Boltzmann equations with Maxwell's kinetic boundary conditions to derive non-local hydrodynamics with arbitrary accommodation. Focusing on shear-mode dynamics, we obtain explicit steady state solutions in terms of Fourier integrals and closed-form expressions for the mean flow and the stress. We demonstrate that the exact non-local fluid model correctly predicts several rarefaction effects with accommodation, including the Couette flow and thermal creep in a plane channel.  
\end{abstract}

\maketitle

\noindent 
Exact hydrodynamics have been recently derived in \cite{kogelbauerrarefied2024} for general near-equilibrium Boltzmann equations by projection onto the slow manifold spanned by hydrodynamic eigenvectors. Salient features of these macroscopic field equations are their inherent spatial non-locality, criticality in wave number and the existence of an entropy-dissipation balance. The non-local hydrodynamics give a solution to Hilbert's sixth problem \cite{gorban2014hilbert} on the linear level without any smallness assumption on Knudsen number and are equivalent to the full summation of the Chapman--Enskog series \cite{gorban1996short,kogelbauer2020slow}. The exact non-local hydrodynamics thus provide a dynamically optimal description of a rarefied gas in terms of conventional macroscopic fields (density, momentum and energy) that goes way beyond the Navier--Stokes equation. It is well known that characteristic properties of rarefied gas flows such as the temperature jump or the velocity slip at the wall, cannot be captured by conventional hydrodynamics \cite{hadjiconstantinou2006limits}. Analogously, the interaction of the rarefied flow with the reflective-diffusive boundary, leading to the Knudsen layer, requires higher-order hydrodynamic models \cite{struchtrup2007h} or kinetic simulations \cite{ansumali2007}. 

In this Letter, we extend the exact non-local fluid equations by incorporating kinetic boundary conditions. We provide a general solution method and demonstrate that criticality is the dominant feature responsible for rarefaction effects. While our methodology applies to the full non-local hydrodynamics, we demonstrate it by focusing on problems which only include the shear mode, thus allowing for an effectively one-dimensional kinetic modeling. The non-local hydrodynamics with accommodation are exemplified on two classical rarefaction problems, planar Couette flow and thermal creep, for which tabulated solutions for the stresses and flow rates have been calculated for the Bhatnagar--Gross--Krook (BGK) collision model \cite{loyalka1979some,barichello2001unified}.

Consider the non-local hydrodynamics in frequency space,
\begin{equation}\label{equ}
    \partial_t \hat{u}(k) = \hat{\lambda}(k) \hat{u}(k),
\end{equation}
where $\hat{u}$ is the spatial Fourier transform of the velocity field $u$ and $\hat{\lambda}(k)$ is the frequency-dependent shear-type mode of the underlying Boltzmann operator, acting as a Fourier multiplier \cite{kogelbauerrarefied2024,kogelbauer2024exact}. The Fourier multiplier satisfies the asymptotics
\begin{equation}\label{Ok2}
    \hat{\lambda}(k)= \mathcal{O}(k^2),\quad k\to 0,
\end{equation}
corresponding to the Navier--Stokes limit in consistency with the Chapman--Enskog expansion, while it remains bounded for $k\to\infty$. Indeed, there exists a critical wave number $k_{\rm crit}$ such that $\hat{\lambda}(k)$ exists as an isolated eigenvalue only for $|k|<k_{\rm crit}$, see \cite{kogelbauerrarefied2024}, and we set 
\begin{equation}\label{ext}
    \hat{\lambda}(k) = -\nu,\quad |k|>k_{\rm crit},
\end{equation}
as an extension of the shear-type mode in frequency space, where $\nu$ is the collision frequency of the underlying kinetic model evaluated at $\bm{v}=0$, see \cite{cercignani1988boltzmann}. The extension of the hydrodynamic mode \eqref{ext} can thus be interpreted as a way to include information about the essential spectrum of the Boltzmann kinetic equation into the non-local hydrodynamics.

Equation \eqref{equ} is based on an underlying kinetic operator of Boltzmann type in frequency space, $\mathcal{L}_{\bm{k}}=-\ri\bm{k}\cdot \bm{v} + \mathcal{Q}$, where $\mathcal{Q}$ is a self-adjoint collision operator, see \cite{kogelbauerrarefied2024}. In this work, we restrict our analysis to one spatial dimension and assume that the velocity field is aligned with the shear flow direction. Taking problem-specific moments in velocities then leads to a one-dimensional kinetic operator in frequency space of the form $L_k=-\ri k v+\mathcal{Q}_s$ and a one-dimensional distribution function $f(x,v)$, where $\mathcal{Q}_s$ is the self-adjoint projected collision operator, in the shear direction \cite{williams2001review}. The non-local evolution equation \eqref{equ} is then derived from the projection onto the hydrodynamic slow manifold,
\begin{equation}\label{fhydro}
    f_{\rm hydro}(x, v) = \frac{1}{\sqrt{2\pi}}\int dk\,  \hat{f}_{\lambda}(v,k) \hat{u}(k)  e^{\ri k x}
\end{equation}
where  $\hat{f}_\lambda$ is the eigenfunction of $L_{k}$ with eigenvalue $\hat{\lambda}$, indexed by wave-number.

Below, we use units such that the thermal speed is $v_T = \sqrt{k_BT/m}=1$ and
the global Maxwellian is  $f^{\rm eq}(v) = (1/\sqrt{2\pi})\exp(-v^2/2)$. 
The general reflective-diffusive boundary conditions \cite{cercignani1988boltzmann} for a one-dimensional kinetic equation on the interval of length $L$ take the form,
\begin{equation}\label{kineticBC}
    \begin{cases}
        f(0,v) - (1-\alpha)f(0,-v)  & =  \phi_0(v,\alpha),\quad v>0,\\
        f(L,-v) - (1-\alpha)f(L,v) & =   \phi_L(v,\alpha) ,\quad v>0,
    \end{cases}
\end{equation}
where $L$ is the distance between the walls of the channel, $0\leq \alpha\leq 1$ is the accommodation coefficient, and the polynomials $\phi_{0,L}$ are derived from the equilibrium distributions at the respective walls by taking velocity moments accordingly \cite{williams2001review}.

To derive a steady state solution to \eqref{equ} between the two plates, we first observe that any information that defines a non-trivial steady state has to be confined to the boundary points. A classical result in distribution theory \cite{hörmander2015analysis} implies that the boundary data has to be a superposition of two delta distributions and their derivatives.   
We will consider only the leading order contribution of these boundary functions, thus assuming two delta functions of opposing sign at the respective boundaries, which leads to the equation,
\begin{equation}\label{eqdelta}
    \int_{-\infty}^\infty dy\, \int_{-\infty}^\infty dk\, u(x-y)\hat{\lambda}(k) e^{\ri k x} =\overline{\sigma} \Big(\delta(x-L)-\delta(x)\Big),
\end{equation}
where $\delta$ is the Dirac delta function and where the connection of the parameter $\overline{\sigma}$ to the stress component will become apparent later. Equation \eqref{eqdelta} can now readily be solved in Fourier space via division by $\hat{\lambda}$ and addition of an integration constant.
The explicit steady state solution to equation \eqref{equ} and \eqref{eqdelta} respectively on the interval $(0,L)$ is then given by,
\begin{equation}\label{generalsteady}
    u(x) = \overline{u} + \frac{\overline{\sigma}}{2\pi} \int_{-\infty}^\infty dk\,\frac{(e^{-\ri L k}-1)}{\hat{\lambda}(k)} e^{\ri k x} ,
\end{equation}
for the mean flow component $\overline{u}$ and the constant stress component $\overline{\sigma}$.

The integral expression in \eqref{generalsteady} has to be interpreted as a principal value for small wave numbers, whose convergence is guaranteed by the asymptotics \eqref{Ok2}. Indeed, the choice of delta functions with opposing sign guarantees that the numerator in \eqref{generalsteady} is of order $k$ thus implying that the integrand is of order $k^{-1}$. A contour-integration argument then shows that \eqref{generalsteady} exists for $k$ small enough. For large wave numbers,  the integral does not converge in the classical sense as the inverse Fourier transform of a bounded signal and has to be interpreted as a generalized function. We can thus rewrite \eqref{generalsteady} as 
\begin{equation}\label{uexplicit}
\begin{split}
         u(x) & = \overline{u} + \frac{\overline{\sigma}}{2\pi} \int_{-k_{\rm crit}}^{k_{\rm crit}} dk\,\left(\frac{1}{\hat{\lambda}(k)}+\frac{1}{\nu}\right)(e^{-\ri L k}-1)e^{\ri k x} \\
         &+\frac{\overline{\sigma}}{2\pi\nu} \Big(\delta(x)-\delta(x-L)\Big),
\end{split}
\end{equation}
We emphasize the inherently weak character of the steady state solution to \eqref{equ}, which splits into a smooth part, the first two terms in \eqref{uexplicit}, and a distributional part due to the boundedness of $\hat{\lambda}$ by \eqref{ext}. In the limit $\rm{Kn}\to 0$, however, solution \eqref{generalsteady} converges to the classical steady-state solution of the heat equation, since $\nu,k_{\rm crit}\to\infty$, thus proving consistency with the Chapman--Enskog expansion. 

While the Fourier multiplier $\hat{\lambda}$ is defined for all frequencies and hence on the whole real line, solution \eqref{generalsteady} only uses information inside the interval $(0,L)$ and at its boundary. Indeed, the steady state solution \eqref{uexplicit} becomes singular at the boundary, while its smooth component can be extended to the real line. This phenomenon is reminiscent of the tunneling effect in quantum mechanics \cite{razavy2013quantum}. 
We stress that all these properties are consequences of the non-local character of \eqref{equ}.
Furthermore, on the slow manifold, the stress is related to the velocity field via the Fourier multiplier \cite{kogelbauerrarefied2024},
\begin{equation}\label{sigmak}
    \hat{\sigma}(k) = -\frac{\hat{\lambda}(k)}{\ri k } \hat{u}(k).
\end{equation}
From the asymptotics \eqref{Ok2}, we necessarily have that $\hat{\sigma}(k) = \mathcal{O}(k)$, while the application of \eqref{sigmak} to the explicit solution \eqref{generalsteady} implies that
\begin{equation}
    \sigma(x) = \overline{\sigma},\ 0<x<L. 
\end{equation}

Thus, the steady state solution is defined by \eqref{uexplicit} up to the two constants, the mean flow $\overline{u}$ and the stress $\overline{\sigma}$.
The latter are derived from the kinetic boundary conditions \eqref{kineticBC}, where we have to balance the corresponding first-order fluxes on the hydrodynamic slow manifold \cite{grad1949kinetic}.
To that end, we multiply both equations in \eqref{kineticBC} with $v$ and integrate over non-negative velocities.
\begin{figure}[t!]
    \centering
    \includegraphics[width=0.9\linewidth]{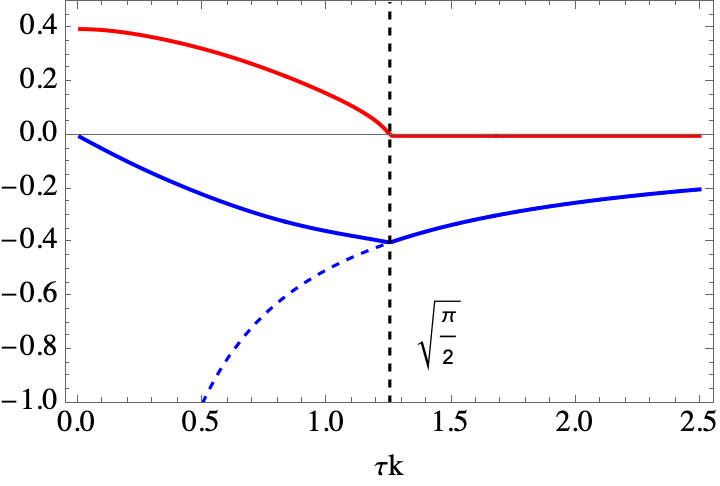}
    \caption{Real part (red) and imaginary part (blue) of the Fourier multiplier $\hat{s}$  for the BGK model  \eqref{eq:sBGK}. For $k\tau>\sqrt{\frac{\pi}{2}}$ (critical wave number, dashed vertical line), the real part vanishes while the imaginary part becomes $-\frac{1}{2 k\tau}\ri$ (dashed blue line).}
    \label{figs}
\end{figure}
\noindent
Restricting the distribution function appearing on the left-hand side of \eqref{kineticBC} to the hydrodynamic slow manifold \eqref{fhydro} and taking half-space fluxes accordingly leads to the Fourier multiplier, 
\begin{equation}\label{defs}
        \hat{s}(k) = \frac{1}{\sqrt{2\pi}} \int_{0}^{\infty} dv\, e^{-\frac{v^2}{2}}  v \hat{f}_{\lambda}(v,k).
\end{equation}
The multiplier \eqref{defs} satisfies the symmetry,
\begin{equation}
\hat{s}(-k)=\hat{s}^*(k), 
\end{equation}
which follows from the general symmetry of the shear-mode eigenvectors \cite{kogelbauer2024exact,kogelbauerrarefied2024},
\begin{equation}
\label{symf}
\hat{f}_{\lambda}(v,-k)=\hat{f}_{\lambda}^*(v,k).
\end{equation}
Similarly, by taking half-spaces fluxes for the boundary functions on the right-hand side of \eqref{kineticBC}, we obtain
\begin{equation}\label{defsigmaj}
    \sigma_{0,L}(\alpha) = \frac{1}{\sqrt{2\pi}}\int_0^\infty  dv e^{-\frac{v^2}{2}} v\phi_{0,L}(v,\alpha).
\end{equation} \\
Combining \eqref{defs} with \eqref{defsigmaj}, we arrive at a linear system for the mean flow $\overline{u}$ and the flux $\overline{\sigma}$,
\begin{equation}\label{matrixeq}
    \begin{pmatrix}
        \alpha & \eta(L)\\
        \alpha & -\eta(-L)
    \end{pmatrix} \left(\begin{array}{c}
         \overline{u}\\
          \overline{\sigma} 
    \end{array}\right) = 
    \left(\begin{array}{c}
         \sigma_{0}\\
          \sigma_{L} 
    \end{array}\right),
\end{equation}
where we have defined the coefficient function
\begin{equation}\label{defeta}
    \eta(L) = 2\Re\int_{0}^\infty dk \, \frac{(1-e^{-\ri L k})}{\hat{\lambda}(k)}[\hat{s}(k)-(1-\alpha)\hat{s}^*(k)],
\end{equation}
and have omitted  the dependence on $\alpha$ to ease notation.

System \eqref{matrixeq} has a unique solution provided that $\alpha\neq 0$. 
Indeed, it is well known that for purely reflective boundaries ($\alpha = 0$), the mean-flow component cannot be recovered uniquely from the boundary conditions \cite{loyalka1971thermal}.
This property is also reflected on the level of the hydrodynamic manifold \eqref{fhydro}, which satisfies the symmetry \eqref{symf}.
The solution to \eqref{matrixeq} is given explicitly by
\begin{equation}\label{solusigma}
    \overline{u}= \frac{\sigma_{0}\eta(-L)+\sigma_{L}\eta(L)}{\alpha[\eta(-L)+\eta(L)]},\quad \overline{\sigma}= \frac{\sigma_{0}-\sigma_{L}}{\eta(-L)+\eta(L)},
\end{equation}
whenever $\alpha\neq 0$, thus expressing the mean flow and the flux in terms of the width of the channel $L$ and Maxwell's accommodation coefficient $\alpha$. Thus, the solution to the steady-state problem is reduced to quadratures \eqref{defeta} for a generic Boltzmann equation.

To assess the accuracy of the non-local hydrodynamics, we compare \eqref{solusigma} to numerical solutions of the Bhatnagar--Gross--Krook (BGK) kinetic equation \cite{bhatnagar1954model}. For a complete and explicit spectral theory of the linear BGK operator, we refer to \cite{kogelbauer2024exact}:
The eigenfunction associated to the shear mode takes the form,
\begin{equation}
    f_\lambda(v,k) = \frac{1}{1+\ri \tau kv + \tau\hat{\lambda}(k)},
\end{equation}
where $\tau=1/\nu$ is the  relaxation time of the BGK equation. Furthermore, the BGK shear mode $\hat{\lambda}$ is the solution to the equation \cite{kogelbauer2024exact},
\begin{equation}
    Z\left( \ri \frac{\tau \hat{\lambda}+1}{\tau k} \right) = \ri \tau k,
\end{equation}
where $Z(\zeta) = Z^+(\zeta)-Z^{+}(-\zeta)$ is the plasma dispersion function and
\begin{equation}\label{eq:Zfunction}
Z^+(\zeta) = \frac{1}{\sqrt{2\pi}}\int_{0}^\infty dv \frac{e^{-\frac{v^2}{2}}}{v-\zeta},
\end{equation}
is the incomplete plasma dispersion function \cite{baalrud2013incomplete}, while the critical wave number of the shear mode is,
\begin{equation}
    k_{\rm crit} = \sqrt{\frac{\pi}{2}}\frac{1}{\tau}.
\end{equation}
We consider equation \eqref{equ} on the interval $x\in (0,L)$ and define the Knudsen number as
\begin{equation}
    {\rm Kn} = \frac{\tau}{L},
\end{equation}
recalling that we have chosen units such that $v_{T} =1$. 
For the BGK shear mode, the Fourier multiplier \eqref{defs} takes explicit form,
\begin{equation}\label{eq:sBGK}
    \hat{s} = \frac{1}{2\ri k \tau} + \frac{\tau\hat{\lambda}+1}{\tau^2 k^2} Z^+\left( \ri\frac{\tau\hat{\lambda}+1}{\tau k} \right).
\end{equation}
Function \eqref{eq:sBGK} is shown in Fig.\ \ref{figs}.\\
\begin{figure}[t!]
    \centering
    \includegraphics[width=0.9\linewidth]{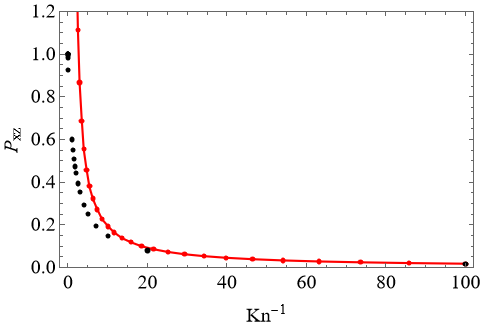}
    \caption{Normalized stress for the planar Couette flow ($\tau=1$, $\alpha=1$). Solid red line with markers: present rarefied hydrodynamics; Black markers: tabulated numerical calculation for the BGK model \cite{barichello2001unified}.}
    \label{FigCouette}
\end{figure}
First we consider the rarefied planar Couette flow generated by the viscous forces of two parallel plates moving in opposite directions. In this case \cite{barichello2001unified},  
\begin{equation}
    \phi_{0}(v,\alpha) = \alpha, \quad  \phi_{L}(v,\alpha) = -\alpha.
\end{equation}
The normalized stress \cite{williams2001review} is derived from the explicit $\overline{\sigma}$-solution \eqref{solusigma} as 
\begin{equation}\label{pxz}
    P_{\rm xz} =  \frac{2\pi\alpha}{\sqrt{2}[\eta(L)-\eta(-L)]}.
\end{equation}
\begin{figure}[h!]
    \centering
    \includegraphics[width=0.9\linewidth]{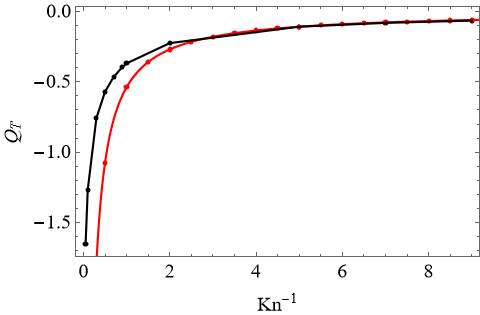}
    \caption{Normalized flow rate for the thermal creep flow ($\tau=1$, $\alpha=0.5$). Solid red line with markers: present rarefied hydrodynamics; Black markers: tabulated numerical calculation for the BGK model \cite{barichello2001unified}. }
    \label{FigThermalCreep}
\end{figure}
\noindent
Fig.\ \ref{FigCouette} shows a comparison of \eqref{pxz} to numerical results of the full BGK kinetic equation, indicating excellent agreement up to Knudsen numbers $\rm Kn \sim \mathcal{O} (10)$. 
The coefficient function \eqref{defeta} satisfies the asymptotics
\begin{equation}\label{asymptoticeta}
\begin{split}
\eta(L) & = \left(\frac{\hat{s}(0)}{\tau}\int_{0}^{\infty} dy\, \frac{1-e^{-\ri y}}{y^2}\right)L + \mathcal{O}(L^2)\\
 & = \sqrt{2\pi}\frac{L}{\tau} +\mathcal{O}(L^2),\quad L\to \infty,
\end{split}
\end{equation}
where we have used the asymptotics $\hat{\lambda}(k) = - \tau k^2+\mathcal{O}(k^4)$, $k\to 0$, of the BGK shear mode together with $\hat{s}(0)=1/\sqrt{2\pi}$. Combining \eqref{asymptoticeta} with \eqref{pxz} gives
\begin{equation}
    P_{\rm xz} = \sqrt{\pi}\frac{\tau}{L}+\mathcal{O}(L^{-2}),\quad L\to \infty, 
\end{equation}
which is consistent with the Navier--Stokes asymptotics of the planar Couette flow, see \cite{loyalka1979some}.\\
As a second example, we consider the thermal creep problem, for which 
\begin{equation}
  \phi_{0}(v,\alpha) = \phi_{L}(v,\alpha)  = \frac{\alpha}{2}\left(v^2-\frac{1}{2}\right),
\end{equation}
see \cite{barichello2001unified}. The macroscopic quantity of interest is the normalized flow rate,
\begin{equation}
    Q_T(L) = -\frac{2\tau}{L^2}\int_0^Ldx\,  u(x).
\end{equation}
The explicit solution \eqref{solusigma} predicts a vanishing flux $\overline{\sigma}$, while it recovers the mean flow component as
\begin{equation}
    \overline{u} =  \frac{1}{4\sqrt{2\pi}}\int_{0}^\infty dv\, v(2v^2-1)e^{-\frac{v^2}{2}} = \frac{3}{4\sqrt{2\pi}},
\end{equation}
and the normalized flow rate as
\begin{equation}\label{QT}
    Q_{T}(L) = -\frac{2\tau}{L} \overline{u}= -\frac{3}{2\sqrt{2\pi}}\frac{\tau}{L}. 
\end{equation}
\noindent
Figure \ref{FigThermalCreep} shows a comparison of \eqref{QT} to numerical results of the full BGK kinetic equation, indicating excellent agreement up to Knudsen numbers $\rm Kn \sim\mathcal{O} (1)$.

In summary, we presented an extension of the exact rarefied hydrodynamics \cite{kogelbauerrarefied2024} by general kinetic boundary conditions. We provided a closed-form solution to a class of rarefied flow problems with general accommodation and showed a quantitative comparison for the planar Couette flow and the thermal creep, giving excellent agreement over a remarkably wide range of Knudsen numbers. Thus, the practical solution to Hilbert's sixth problem \cite{gorban2014hilbert} on the linear level has been extended to include boundary conditions as well. We demonstrated that rarefaction effects are solely a consequence of the non-locality of exact hydrodynamics, a feature which is not shared by any approximate hydrodynamics such as Navier--Stokes or Burnett \cite{chapman1990mathematical}.
Finally, while we focused on the shear mode dynamics in this work, the above approach is applicable to the general non-local hydrodynamics for the full set of modes \cite{kogelbauerrarefied2024}.\\
\begin{acknowledgments}
This work was supported by European Research Council (ERC) Advanced Grant  834763-PonD and
Swiss National Science Foundation (SNSF) grant No. 200021-228065.
Computational resources at the Swiss National  Super  Computing  Center  CSCS  were  provided  under the grant s1286.
\end{acknowledgments}

\bibliographystyle{apsrev4-1}
\bibliography{Knudsen}

\end{document}